\documentclass[a4paper]{amsart}
\pdfoutput=1
\usepackage[english]{babel}
\expandafter\let\csname equation*\endcsname\relax

\expandafter\let\csname endequation*\endcsname\relax
\usepackage{amsmath}
\usepackage{amssymb}
\usepackage{hyperref}
\usepackage{amsfonts}
\usepackage{tikz}
\usepackage{comment}

          \newcommand\hlight[1]{\tikz[overlay, remember picture,baseline=-\the\dimexpr\fontdimen22\textfont2\relax]\node[rectangle,fill=white!50,rounded corners,fill opacity = 0.2,draw,thick,text opacity =1] {$#1$};}

\usepackage[latin9]{inputenc}
\usepackage{graphicx}
\usepackage{lipsum,parskip,booktabs,graphicx,tabulary,tabularx}
\usepackage{dcolumn}
\usepackage{bm}
\usepackage{physics}
\usepackage{mathtools}
\usepackage{amsthm}
\usepackage{dsfont}
\usepackage{tikz}
\usepackage{dashrule}
\usepackage{arydshln}
\usepackage{array}
\usepackage{academicons}
\newcommand{\orcid}[1]{\href{https://orcid.org/#1}{\textcolor[HTML]{A6CE39}{\aiOrcid}}}
\usetikzlibrary{matrix}

\newlength\mylen
\usepackage{array} 
\newcolumntype{C}{>{\hfil$}p{\mylen}<{$\hfil}} 

\newtheorem{thm}{Theorem}[section]

\newtheorem{prop}[thm]{Proposition}

\theoremstyle{definition}
 
\newtheorem{exam}[thm]{Example} 

\theoremstyle{remark}
\newtheorem{rem}[thm]{Remark}

\numberwithin{equation}{section}

\newcommand{\cH}{{\mathcal H}}

\newcommand{\bC}{{\mathbb C}}

\newcommand{\cM}{{\mathbb M}}

\newcommand{\ba}{\begin{array}}
\newcommand{\ea}{\end{array}}
\newcommand{\be}{\begin{eqnarray*}}
\newcommand{\ee}{\end{eqnarray*}}
\newcommand{\beg}{\begin{eqnarray}}
\newcommand{\eeg}{\end{eqnarray}}
\newcommand{\beq}{\begin{equation}}
\newcommand{\eeq}{\end{equation}}

\begin{document}
\title{Characterization of $k$-positive maps}

\author{Tomasz M{\l}ynik}
\address{
Institute of Theoretical Physics and Astrophysics, Faculty of Mathematics, Physics, and Informatics, University of Gda\'nsk, Wita Stwosza 57, 80-308 Gda\'nsk, Poland}
\email{tomasz.mlynik@phdstud.ug.edu.pl}

\author{Hiroyuki Osaka}
\address{
Department of Mathematical Sciences, Ritsumeikan University, Kusatsu, Shiga 525-8577, Japan}
\email{osaka@se.ritsumei.ac.jp}

\author{Marcin Marciniak}
\address{Institute of Theoretical Physics and Astrophysics, Faculty of Mathematics, Physics, and Informatics, University of Gda\'nsk, Wita Stwosza 57, 80-308 Gda\'nsk, Poland}
\email{marcin.marciniak@ug.edu.pl}

\keywords{positive maps, k-positivity, Choi matrix}

\begin{abstract}
We present a general characterization of k-positivity for a positive map in terms of the estimation of the Ky Fan norm of the matrix constructed from the Kraus operators of the associated completely positive map. Combining this with the result given by Takasaki and Tomiyama we construct a family of positive maps between matrix algebras of different dimensions depending on a parameter. The estimate bounds on the parameter to obtain the $k$-positivity are better than those derived from the spectral conditions considered by Chru\'sci\'nski and Kossakowski. We further look with special attention at the case where we give the precise bound for the regions of decomposability.
\end{abstract}

\maketitle

\section{\label{sec:level1}Introduction}
Examining the mixed states of composite quantum systems to determine the presence of quantum correlation, or whether the state is entangled or separable \cite{QuantCorr,QEnt}, is a fundamental issue in quantum information theory. This is particularly important for private and quantum communication problems \cite{QuantComm}. Most communication-related experiments are conducted based on the entanglement between qubits. For low-dimensional complex systems, such as qubit-qubit and qubit-qutrit, the Peres-Horodecki criterion \cite{hhh} fully characterizes separable states, stating that a state is separable if and only if its partial transpose is positive. However, technological advancements now allow for the experimental control of composite systems with a larger number of degrees of freedom \cite{DLBPA,Giov,FLHLPZ,Bav}. For high-dimensional systems, there is no exhaustive separability condition. 

The aim of our paper is to explore the potential for constructing indecomposable positive $k$ maps for higher $k$ values.
We investigate a family of linear maps $\Phi_a$ acting from $\mathbb{M}_m$ to $\mathbb{M}_n$ ($m \leq n$), indexed by a positive parameter $a$. Our construction is inspired by \cite{tt2} and represents a specific instance of the general construction outlined in \cite{chk}. Using different approaches from those in \cite{chk}, we estimate the bounds of the parameter $a$ that ensure the $k$-positivity of $\Phi_a$. These estimates prove to be significantly better than those derived from the spectral condition considered by Chru\'sci\'nski and Kossakowski. We further focus on the scenario where $n$ and $m$ differ by one, providing an explicit analytical formula for the constraints on the parameter $a$ that guarantees $k$-positivity.
We also pay special attention to cases where we establish the precise bounds for the regions of decomposability. Furthermore, we demonstrate that for sufficiently large differences $n-m$, some maps are neither completely positive nor completely copositive within the region of $k$-positivity for higher $k$.

\section{Notation and preliminaries}

By $\mathbb{M}_n$ we denote the algebra of all square $n\times n$-matrices with complex coefficients and by $\mathbb{M}_n^+$ the cone of positive semidefinite matrices from $\mathbb{M}_n$. 
Given $m,n\in\mathbb{N}$, a linear map $\phi:\mathbb{M}_m\to\mathbb{M}_n$ is called a \textit{positive map} if $\phi(\mathbb{M}_m^+)\subset\mathbb{M}_n^+$. 
The identity map and the transpose map on $\mathbb{M}_n$ are denoted by $\mathrm{id}_n$ and $\tau_n$, respectively. For $k\in\mathbb{N}$, a map $\phi$ is called \textit{$k$-positive} (respectively, \textit{$k$-copositive}) if the ampliation map $\mathrm{id}_k\otimes \phi : \mathbb{M}_k\otimes\mathbb{M}_m\to\mathbb{M}_k\otimes\mathbb{M}_n$ (respectively, $\tau_k\otimes\phi :\mathbb{M}_k\otimes\mathbb{M}_m\to\mathbb{M}_k\otimes\mathbb{M}_n$) is positive. 
We say that a map $\phi$ is \textit{completely positive} (respectively \textit{completely copositive}) if it is $k$-positive (respectively $k$-copositive) for every $k\in\mathbb{N}$. A positive map $\phi$ is said to be \textit{decomposable} if it is a sum of a completely positive map and a completely copositive one. The sets of all linear, positive, $k$-positive, $k$-copositive, completely positive, completely copositive and decomposable maps from $\cM_m$ into $\cM_n$ will be denoted by $L(\cM_m,\cM_n)$, $\mathbb{P}(m,n)$, $\mathbb{P}_k(m,n)$, $\mathbb{P}^k(m,n)$, $\mathbb{P}_\infty(m,n)$, $\mathbb{P}^\infty(m,n)$ and $\mathbb{D}(m,n)$, respectively. We will omit the dimensions $m,n$ in case it does not lead to confusion.

Let $\cH_1$, $\cH_2$ be finite dimensional Hilbert spaces with $d_i=\dim\cH_i$, $i=1,2$, and let $A:\cH_1\to\cH_2$ be a linear operator. 
For $k=1,2,\ldots,d$, where $d=\min\{d_1,d_2\}$, we define Ky Fan $k$-norm of $A$ \cite{kyfan}
\begin{equation}
\Vert A \Vert_{(k)} =\sum_{i=1}^k s_i(A), 
\label{e:kyfan}
\end{equation}
where $s_1(A) \geq\ldots\geq s_d (A)$ are the singular values of $A$. Clearly, for
$k = 1$ one recovers the operator norm $\Vert A\Vert_{(1)}= \Vert A \Vert$ and if $d_1 = d_2 = d$, then for $k = d$
one reproduces the trace norm $ \Vert A \Vert_{(d)} = \Vert A \Vert_\mathrm{tr}:=\Tr((A^\dagger A)^{1/2})$. 
Note that a Ky Fan $k$-norm may be equivalently introduced as follows. By $\mathrm{Proj}_k(\cH)$ we denote the set of all orthogonal projections on a Hilbert space $\cH$ with trace (i.e. dimension of the image) equal to $k$.
It is easy to show that
\begin{eqnarray*}
\Vert A \Vert_{(k)}=
\max
\left\{\Tr (Q(AA^\dagger)^\frac{1}{2}):\,Q\in\mathrm{Proj}_k(\cH_2)\right\}.
\end{eqnarray*}

\section{Characterisation of $k$-positivity}
Let $\mathbf{Tr}:\mathbb{M}_m\to\mathbb{M}_n$ be a map defined by $\mathbf{Tr}(X)=\Tr(X)\mathds{1}_n$ for $X\in\mathbb{M}_m$.
Each positive map is a scalar multiple of a map of the form $\mathbf{Tr} - \psi$ with $\psi$ being a uniquely determined completely positive map. Since every completely positive map has its Kraus representation, we arrive at the following general form of a positive map $\phi:\mathbb{M}_m\to\mathbb{M}_n$
\begin{equation}
\phi(X) = a\Tr(X)\mathds{1}_n - \sum_{\alpha=1}^N K_\alpha X K_\alpha^\dagger,\qquad X\in\mathbb{M}_m,
\label{eqn:1}
\end{equation}
where
$K_\alpha:\mathbb{C}^m\to\mathbb{C}^n$ for $\alpha=1,2,\ldots,N$, are linear operators and $a$ is a positive number.
Assume that the operators $K_\alpha$ are fixed. Our aim is to characterize such a range of numbers $a$ that the map $\phi$ given by \eqref{eqn:1} is $k$-positive for $k\in\mathbb{N}$. 

Let us observe that the Choi matrix $C_a$ of $\phi$ has the form $C_a=a\mathds{1}_m\otimes\mathds{1}_n-C$, where
$$C=\sum_{i,j=1}^m e_{ij}\otimes \sum_{\alpha=1}^N K_\alpha e_{ij}K_\alpha^\dagger$$
We define 
\begin{equation}
a_k=
\sup\{\norm{(\mathds{1}_m\otimes Q)C(\mathds{1}_m\otimes Q)}:\,
Q\in\mathrm{Proj}_{k}(\bC^n)\}.
\label{e:muk}
\end{equation}

\begin{prop}\label{prp:k-positivity}
The map $\phi$ defined in \eqref{eqn:1} is $k$-positive if and only if $a\geq a_k$.
\end{prop} 
\begin{proof}
Assume $a\geq a_k$. According to \cite[Prop. 1.1]{tt} it is enough to show that $(\mathds{1}_m\otimes Q)C_{a}(\mathds{1}_m\otimes Q)$ is positive for every $Q\in\mathrm{Proj}_k(\mathbb{C}^n)$. Let $\ket{\xi}\in\mathbb{C}^m\otimes\mathbb{C}^n$ be such that $\norm{\xi}=1$. 
We can assume that $\ket{\xi}\in(\mathds{1}_m\otimes Q)(\bC^m\otimes\bC^n)$. Then 
\begin{eqnarray*}
\lefteqn{
\bra{\xi}(\mathds{1}_m\otimes Q)C_{a}(\mathds{1}_m\otimes Q)\ket{\xi}
=} \\
&=&a \norm{(\mathds{1}_m\otimes Q)\xi}^2-\bra{\xi}(\mathds{1}_m\otimes Q)C (\mathds{1}_m\otimes Q)\ket{\xi}
\\
&\geq&
a_k - \norm{(\mathds{1}_m\otimes Q)C (\mathds{1}_m\otimes Q)})\geq 0,
\end{eqnarray*}
where the last inequality follows from \eqref{e:muk}.

Now, assume that $a<a_k$. It follows from the definition of supremum that there exists a projection $Q\in\mathrm{Proj}_{k}(\mathbb{C}^n)$ and a unit vector $\ket{\xi}\in (\mathds{1}_m\otimes Q)(\bC^m\otimes\bC^n)$ such that $a<\bra{\xi}(\mathds{1}_m\otimes Q)C (\mathds{1}_m\otimes Q)\ket{\xi}$. Hence
\begin{eqnarray*}
\lefteqn{\bra{\xi}(\mathds{1}_m\otimes Q)C_{a}(\mathds{1}_m\otimes Q)\ket{\xi}=} \\
&=& a-\bra{\xi}(\mathds{1}_m\otimes Q)C (\mathds{1}_m\otimes Q)\ket{\xi}<0
\end{eqnarray*}
what shows that $\phi$ is not $k$-positive.
\end{proof}

We are now ready to formulate our main result specifying the lower bound $a_k$. 
\begin{thm}\label{thm1}
Assume that $\phi$ is of the form (\ref{eqn:1}). It is $k$-positive if and only if 
\begin{equation}
\label{e:Kynorm}
a\geq \sup_{\zeta}\norm{\left( \sum_{\alpha=1}^N{\zeta_\alpha}K_\alpha\right)\left(\sum_{\beta=1}^N\zeta_\beta K_\beta\right)^\dagger}_{(k)},
\end{equation}
where the supremum is taken over all vectors $\ket{\zeta} = (\zeta_1, \zeta_2,\dots, \zeta_{N})^T \in \mathbb{C}^{N}$, such that $\norm{ \zeta }=1$ and $\norm{\cdot}_{(k)}$ denotes the $k$-th Ky Fan norm.
\end{thm}

\begin{proof}
We will show that $a_k$ defined in \eqref{e:muk} is equal to the expression on the right-hand side of \eqref{e:Kynorm}. Let us note that  $C = \sum_{i,j=1}^m e_{ij}\otimes \sum_{\alpha=1}^N K_\alpha e_{ij}K^\dagger_\alpha \in \mathbb{M}_m\otimes\mathbb{M}_n\subset \mathbb{M}_m\otimes\mathbb{M}_{N}\otimes\mathbb{M}_n \simeq \mathbb{M}_m\otimes \mathbb{M}_{N}(\mathbb{M}_n)$ where the embedding is of the form
\begin{equation} 
X\otimes Y\mapsto X\otimes 
\left(
\begin{array}{cccc} 
Y & 0 & \cdots & 0 \\ 
0 & 0 & \cdots & 0 \\ 
\vdots & \vdots & \ddots & \vdots \\ 
0 & 0 & \cdots & 0 
\end{array}
\right).
\end{equation}  
This embedding is isometric. Let us define matrices $\mathbf{K}\in \mathbb{M}_{N}(\mathbb{M}_{n,m})$, $\mathbf{e}_{ij}\in\mathbb{M}_N(\mathbb{M}_m)$ and $\mathbf{Q}\in\mathbb{M}_N(\mathbb{M}_n)$ by
$$
\mathbf{K}=\left(\begin{array}{ccc} K_0 & \cdots & K_r \\ 0 & \cdots & 0 \\ \vdots & \ddots & \vdots \\ 0 & \cdots & 0 \end{array}\right),
\quad
\mathbf{e}_{ij}=
\mathds{1}_{N}\otimes e_{ij}=
\left(\begin{array}{cccc} e_{ij}&0&\cdots& 0\\ 0& e_{ij}& \cdots & 0\\ \vdots & \vdots &\ddots &\vdots\\ 0&0& \cdots& e_{ij} \end{array}\right),
$$
$$
\mathbf{Q}=
\mathds{1}_{N}\otimes Q=
\left(\begin{array}{cccc} Q & 0 & \cdots & 0 \\ 0 & Q & \cdots & 0 \\ \vdots & \vdots &\ddots &\vdots\\0&0& \cdots& Q \end{array}\right).
$$
Thus we have

\begin{eqnarray*}
\lefteqn{\norm{(\mathds{1}_m\otimes Q)C(\mathds{1}_m\otimes Q)}=}\\ 
&=&\norm{(\mathds{1}_m\otimes Q)\left(\sum_{i,j=1}^m e_{ij}\otimes\sum_{\alpha=1}^N K_\alpha e_{ij}K_\alpha^\dagger\right)(\mathds{1}_m\otimes Q)} 
\\
&=&
\norm{\sum_{i,j=1}^m e_{ij}\otimes\sum_{\alpha=1}^N QK_\alpha e_{ij}K_\alpha^\dagger Q}
\\
&=&
\norm{\sum_{i,j=1}^m e_{ij}\otimes\left(\begin{array}{cccc}\sum_{\alpha=1}^N QK_\alpha e_{ij}K_\alpha^\dagger Q & 0 & \cdots & 0 \\ 0 & 0 & \cdots & 0 \\ \vdots & \vdots &\ddots &\vdots\\0&0& \cdots& 0 \end{array}\right)  
} 
\\
&=& 
\norm{\sum_{i,j=1}^m e_{ij}\otimes  \mathbf{Q}\mathbf{K}\mathbf{e}_{ij}\mathbf{K}^\dagger\mathbf{Q}} \\
&=&
m\norm{\left(\mathds{1}_m\otimes
\mathbf{Q}\mathbf{K}\right) \left(\frac{1}{m}\sum_{i,j=1}^m e_{ij}\otimes\mathbf{e}_{ij}\right)\left(\mathds{1}_m\otimes
\mathbf{K}^\dagger\mathbf{Q}\right)} .
\end{eqnarray*}
The middle term above is a projection, so 
one obtains
\begin{eqnarray*}
\lefteqn{
\norm{(\mathds{1}_m\otimes Q)C(\mathds{1}_m\otimes Q)}
=}\\
&=&
m\norm{ \left(\frac{1}{m}\sum_{i,j=1}^m e_{ij}\otimes\mathbf{e}_{ij}\right) \left(\mathds{1}_m\otimes \mathbf{K}^\dagger\mathbf{Q}\mathbf{K}\right)  \left(\frac{1}{m}\sum_{i,j=1}^m e_{ij}\otimes\mathbf{e}_{ij}\right) } .
\end{eqnarray*}
Let $\{h_{\alpha\beta}\}_{\alpha,\beta=1}^{N}$ be a system of matrix units in $\mathbb{M}_{N}(\bC)$. Note that $\mathbf{e}_{ij}=\mathds{1}_{N}\otimes e_{ij}$, hence
\begin{eqnarray}
\lefteqn{
\norm{(\mathds{1}_m\otimes Q)C(\mathds{1}_m\otimes Q)}=
}\nonumber\\
&=&
m\left\Vert
\left(\frac{1}{m}\sum_{i,j=1}^m e_{ij}\otimes\mathds{1}_{N}\otimes {e}_{ij}\right) \left(\sum_{\alpha,\beta=1}^{N}\mathds{1}_m\otimes h_{\alpha\beta}\otimes K_\alpha^\dagger Q K_\beta\right)\right. \times \\
&& \times \left.\left(\frac{1}{m}\sum_{i,j=1}^m e_{ij}\otimes\mathds{1}_{N}\otimes {e}_{ij}\right)\right\Vert \nonumber
\\
&=&
m\norm{
\frac{1}{m^2}\sum_{i,j=1}^m\sum_{\alpha,\beta=1}^N e_{ij}\otimes h_{\alpha\beta}\otimes \Tr\left(K_\alpha^\dagger Q K_\beta\right)e_{ij}
}
\nonumber
\\
&=&
\norm{
\left(\frac{1}{m}\sum_{i,j=1}^m e_{ij}\otimes\mathbf{e}_{ij}\right) \left(\mathds{1}_m \otimes \Big(\Tr\left(K_\alpha^\dagger Q K_\beta\right)\Big)_{\alpha,\beta}\otimes \mathds{1}_m\right)
}
\label{e:tr1}
\\
&=& \norm{\Big(\Tr(K_\alpha^\dagger QK_\beta)\Big)_{\alpha,\beta}}.
\label{e:tr2}
\end{eqnarray}
The above matrix $(\Tr(K_\alpha^\dagger QK_\beta))_{\alpha,\beta=1,\ldots,N}$ is a selfadjoint element of $\mathbb{M}_{N}(\bC)$. The last equality between lines \eqref{e:tr1} and \eqref{e:tr2} is due to the fact that the first and the second factors in \eqref{e:tr1} lie in $\mathbb{M}_m\otimes\mathds{1}_{N}\otimes\mathbb{M}_m$ and $\mathds{1}_m\otimes\mathbb{M}_{N}\otimes\mathds{1}_m$ respectively and the operator norm on tensor product is a cross norm. Observe that $\Tr(K_\alpha^\dagger Q K_\beta)= \Tr(K_\alpha^\dagger QQ^\dagger K_\beta)=\Tr(QK_\beta K_\alpha^\dagger Q)$ for every $\alpha,\beta$. 
We have
\begin{eqnarray*}
\lefteqn{\norm{\Big(\Tr(QK_\beta K_\alpha^\dagger Q)\Big)_{\alpha, \beta}}=} \\
&=&
\sup\left\{\left|\sum_{\alpha,\beta=1}^N \overline{\zeta_\alpha}\zeta_\beta \Tr(QK_\beta K_\alpha^\dagger Q)\right|:\,\ket{\zeta}=(\zeta_1,\ldots, \zeta_N)^T\in \mathbb{C}^{N},\,\Vert\zeta\Vert\leq 1\right\}
\\
&=&
\sup_{\zeta}\left|\Tr\left(Q\left(\sum_{\beta=1}^N \zeta_\beta K_\beta\right)\left(\sum_{\alpha=1}^N \zeta_\alpha K_\alpha\right)^\dagger Q\right)\right|.
\end{eqnarray*}

Now according to \eqref{e:muk} we get
\begin{eqnarray*}
a_k &=&  \sup_{Q\in\mathrm{Proj_k}(\mathbb{C}^n)}\;\sup_\zeta \left|\Tr\left(Q\left(\sum_{\beta=1}^N \zeta_\beta K_\beta\right)\left(\sum_{\alpha=1}^N \zeta_\alpha K_\alpha\right)^\dagger Q\right)\right|\\
&=&
\sup_\zeta\;\sup_Q\left|\Tr\left(Q\left(\sum_{\beta=1}^N \zeta_\beta K_\beta\right)\left(\sum_{\alpha=1}^N \zeta_\alpha K_\alpha\right)^\dagger Q\right)\right|\\
&=&
\sup_\zeta\norm{\left(\sum_{\beta=1}^N \zeta_\beta K_\beta\right)\left(\sum_{\alpha=1}^N\zeta_\alpha K_\alpha\right)^\dagger}_{(k)}.
\end{eqnarray*}
This completes the proof.
\end{proof}

\section{Applications}
In this section, we present an explicit construction of a family of $k$-positive maps. 

The main motivation for our consideration are results of \cite{tt2}. This paper describes a linear map acting on $\mathbb{M}_n$ into itself whose Choi matrix is of the form $A - \lambda \ket{\psi_n^+}\!\bra{\psi_n^+}$ where $\lambda \geq 0$,
and $A$ is a positive invertible matrix in $\mathbb{M}_n\otimes\mathbb{M}_n$ acting on the orthogonal complement of $\ket{\psi_n^+}$ in $(\bC^n)^{\otimes 2}$. Their result states that for $k=1,2,\ldots,n-1$, this map is $k$-positive when $A\geq \dfrac{k\lambda}{n-k}\left(\mathds{1}_{n}-\ket{\psi_n^+}\!\bra{\psi_n^+}\right)$ and is not $k$-positive when $\Vert A\Vert<\dfrac{k\lambda}{n-k}$. When applying this result to the case $A=a(\mathds{1}_{n}-\ket{\psi_n^+}\!\bra{\psi_n^+})$, $a>0$, one can deduce that the map 
\begin{equation}\label{e:Phia}\Phi_a(X)=a\Tr(X)\mathds{1}_n-X\end{equation}
is $k$-positive  and is not $(k+1)$-positive if and only if $k\leq a<k+1$.
\begin{rem}
For $a=1$ the map \eqref{e:Phia} is known as a reduction map, while for $a=n-1$ it is nothing but the map considered by Choi in \cite{choii}.
\end{rem} 

\subsection{One-parameter family of maps}
Motivated by \cite{tt2} and \cite{chk} we propose the following construction.
Let $m,n\in\mathbb{N}$ be such that $2\leq m\leq n$. 
Let $\{\ket{e_p}\}_{p=1}^m$ 
and $\{\ket{f_s}\}_{s=1}^n$ 
be standard orthonormal bases in $\bC^m$ and $\bC^n$ respectively.
Consider linear isometries $V_\alpha :\mathbb{C}^m\to\mathbb{C}^n$, $\alpha=0,1,\ldots,r$, where $r=n-m$, given by 
\beq
V_\alpha \ket{e_p}=\ket{f_{p+\alpha}}, \qquad p = 1,2,\ldots, m. 
\eeq 
Given $a>0$, we define a map $\Phi_{a;m,n} :\mathbb{M}_m\to\mathbb{M}_n$, 
\begin{equation}
\Phi_{a;m,n}(X) = a\Tr(X)\mathds{1}_n - \sum_{\alpha=0}^r V_\alpha X V_\alpha^\dagger,
\label{phi_1_parameter}
\end{equation}
for $X\in\mathbb{M}_m$. When dimensions are fixed we will write $\Phi_a$ instead of $\Phi_{a;m,n}$. 

To describe the Choi matrix of $\Phi_a$ 
consider projections $P_\alpha\in\mathbb{M}_m\otimes\mathbb{M}_n$, of the form
\begin{eqnarray}
P_\alpha &=&
\frac{1}{m}\sum_{i,j=1}^m \ket{e_i}\!\bra{e_j}\otimes V_\alpha \ket{e_i}\!\bra{e_j}V_\alpha^\dagger \nonumber\\
&=&
\frac{1}{m}\sum_{i,j=1}^m\ket{e_i}\!\bra{e_j}\otimes \ket{f_{i+\alpha}}\!\bra{f_{j+\alpha}}.
\label{eqn:2}
\end{eqnarray}
Observe that $P_\alpha P_\beta = P_\alpha \delta_{\alpha\beta}$, hence $P=\sum_{\alpha=0}^r P_\alpha$ is a projection too. Therefore the Choi matrix $C_{\Phi_a}$ of the map \eqref{eqn:1} has the form
\begin{equation}
C_a:=C_{\Phi_a} = m(\mu\mathds{1}_m\otimes\mathds{1}_n-P),
\label{eqn:choi}
\end{equation}
where $\mu=a/m$.
\begin{rem}
Observe that for $m=n$ the map $\Phi_{a;n,n}$ is precisely the map given by \eqref{e:Phia}.
\end{rem}

\begin{prop}
\label{w:muk}
If $m,n\in\mathbb{N}$ are such that $2\leq m\leq n$ then $a_k<a_{k+1}$ for $k=1,\ldots,m-1$.
\end{prop}
\begin{proof}
    For $\ket{\zeta}\in\bC^{r+1}$ define $V(\zeta)=\sum_{\alpha=0}^r\zeta_\alpha V_\alpha$. 
Firstly, we will show that $\Vert V(\zeta)V(\zeta)^\dagger\Vert_{(k)}<\Vert V(\zeta)V(\zeta)^\dagger\Vert_{(k+1)}$ for $k=1,2,\ldots,m-1$, when $\Vert \zeta\Vert=1$. Notice that 
$V(\zeta)=(v_{pq}(\zeta))_{1\leq p\leq n;\,1\leq q\leq m}$ where
$$v_{pq}(\zeta)=\begin{cases} \zeta_{p-q}, & \text{if $0\leq p-q\leq m$} \\ 0, & \text{otherwise.} \end{cases}$$
Let $\tilde{\alpha}=\min\{\alpha\in\{0,\ldots,r\}:\,\zeta_\alpha\neq 0\}$. Then the square $m\times m$ submatrix $(v_{p,q}(\zeta))_{\tilde{\alpha}+1\leq p\leq \tilde{\alpha}+m;\,1\leq q\leq m}$ has the following triangular form
\begin{equation}
\left(
\begin{array}{ccccc}
\zeta_{\tilde{\alpha}} & 0 & \cdots & 0 & 0\\
* & \zeta_{\tilde{\alpha}} & \ddots & \ddots & 0   \\
\vdots & \ddots & \ddots & \ddots & \vdots \\
* & \ddots & \ddots   &\zeta_{\tilde{\alpha}} &  0 \\
* &* & \cdots   & * & \zeta_{\tilde{\alpha}}
\end{array}
\right).
\end{equation}
It follows that $K(\zeta)$ has rank $m$, and consequently the matrix $K(\zeta)K(\zeta)^\dagger$ has rank $m$ as well. This in turn implies that the matrix $K(\zeta)K(\zeta)^\dagger$ has $m$ strictly positive eigenvalues, which provides sharp inequalities between its successive Ky Fan norms.  

Now, assume that $a_k=a_{k+1}$ for some $k=1,\ldots,m-1$. Let $\ket {\tilde{\zeta} }\in\bC^{N}$ be  a unit vector such that $a_k=\Vert K(\tilde{\zeta})K(\tilde{\zeta})^\dagger\Vert_{(k)}$. The existence of such a vector follows from the compactness of the set of all unit vectors in $\bC^{N}$. Then we have
$a_{k+1}=a_k=\Vert K(\tilde{\zeta})K(\tilde{\zeta})^\dagger\Vert_{(k)}<\Vert K(\tilde{\zeta})K(\tilde{\zeta})^\dagger\Vert_{(k+1)}$, which leads to a contradiction because $a_{k+1}$ is a supremum of the $k+1$-th Ky Fan norms over all unit vectors $\ket{\zeta}$.
\end{proof}

\subsection{Comment on complete copositivity}

\begin{thm}
Let $n\geq 3m-2$. Then complete copositivity of $\Phi_a$ \eqref{phi_1_parameter} implies $a\geq m$.
\end{thm}
\begin{proof}
Let $\ket{\xi}=\sum_{k=1}^m\ket{e_k}\otimes\ket{f_{2m-k}}$. Then
\begin{eqnarray*}
\lefteqn{C_{a}^\Gamma\ket{\xi}=}\\ 
& = &\left(\sum_{i,j=1}^m\ket{e_i}\!\bra{e_j}\otimes\left(a\delta_{ij}\mathds{1}_n-\sum_{\alpha=0}^r\ket{f_{j+\alpha}}\!\bra{f_{i+\alpha}}\right)\right) \ket{\xi}\\
&=&
\sum_{i=1}^m\ket{e_i}\otimes \sum_{j=1}^m\left(a\delta_{ij}\ket{f_{2m-j}}-\sum_{\alpha=0}^{n-m}\delta_{i+\alpha,2m-j}\ket{f_{j+\alpha}}\right)\\
&=&
\sum_{i=1}^m\left(a-\#S_i\right)\ket{e_i}\otimes\ket{f_{2m-i}},
\end{eqnarray*}
where
$$S_i=\{(j,\alpha)\colon 1\leq j\leq m,\,0\leq \alpha\leq n-m,\,j+\alpha=2m-i\}.$$
If the assumption $n\geq 3m-2$ is satisfied, then $r=n-m\geq 2m-2$, so for every pair $i,j$ ($1\leq i,j\leq m$) there is precisely one number $\alpha\in\{0,1,\ldots,r\}$ such that $j+\alpha=2m-i$. Hence, $\#S_i=m$ for every $i=1,2,\ldots,m$. Consequently, $C_{a}^\Gamma\ket{\xi}=(a-m)\ket{\xi}$. Thus $a-m$ is an eigenvalue of $C_{a}^\Gamma$. If $\Phi_a$ is completely copositive then $C_{a}^\Gamma$ is positive semidefinite, therefore $a-m$ must be non-negative.
\end{proof}

\begin{prop}\label{prp:decomposability}
The map $\Phi_a$ is k-positive and decomposable if $a \geq ma_k \geq r+ 1$.
 \end{prop}
 
 \begin{proof}
 
 From Proposition \ref{prp:k-positivity} $\Phi_a$ is $k$-positive if and only if $a \geq m a_k$.
 
 On the contrary, since 
 \begin{align*}
 C_a &= m(\mu \mathds{1}_m \otimes \mathds{1}_n - P)\\
 &= a\mathds{1}_m \otimes \mathds{1}_n -m P\\
 &\geq ma_k \mathds{1}_m \otimes \mathds{1}_n - mP\\
 &\geq (r + 1)\mathds{1}_m \otimes \mathds{1}_n - m \sum_{\alpha=0}^r P_\alpha\\
 &= \sum_{\alpha=0}^r (\mathds{1}_m \otimes \mathds{1}_n - mP_\alpha).
 \end{align*}
and $(\mathds{1}_m \otimes \mathds{1}_n - mP_\alpha)^\tau$ is positive for all $\alpha$ $(\tau$ is the partial transpose$)$, we know that $\sum_{\alpha=0}^r (\mathds{1}_n \otimes \mathds{1}_m - mP_\alpha)$ is completely copositive.

Hence, since $C_\alpha = \{m(a \mathds{1}_m \otimes \mathds{1}_n - P) - \sum_{\alpha=0}^r(\mathds{1}_m \otimes \mathds{1}_n - mP_\alpha)\} + \sum_{\alpha=0}^r(\mathds{1}_m \otimes \mathds{1}_n - mP_\alpha)$, $\Phi_a$ is decomposable.
\end{proof}

\subsection{Special case $r=1$}
In this section, we consider the case when dimensions $m$ and $n$ differ precisely by $1$.
\begin{thm}\label{thm:r_1}
Let $\Phi_a:\mathbb{M}_m\to\mathbb{M}_{m+1}$ be of the form \eqref{phi_1_parameter} and let $1\leq k\leq m$. $\Phi_a$ is $k$-positive if and only if
\begin{equation}
\label{eqn:thm6}
a\geq k+\sum_{j=1}^k\cos\left(\frac{j\pi}{m+1}\right).
\end{equation}
\end{thm}
\begin{proof}
Let $M=(\sum_{\beta=0}^1\overline{\zeta_\beta}V_\beta)(\sum_{\alpha=0}^1\zeta_\alpha V^\dagger_\alpha)$  be the $(m+1)\times(m+1)$ matrix which appeared in Theorem \ref{thm1}, where $\zeta_0,\zeta_1\in\bC$ such that $|\zeta_0|^2+|\zeta_1|^2=1$. 
It turns out that $M$ is a tridiagonal matrix of the form
\begin{eqnarray*}
\left(\begin{array}{ccccccc}
|\zeta_0|^2 & \overline{\zeta_0}\zeta_1 & 0 & 0 & \cdots & 0 & 0 \\
\zeta_0\overline{\zeta_1} & 1 & \overline{\zeta_0}\zeta_1 & 0 & \cdots & 0 & 0 \\
0 & \zeta_0\overline{\zeta_1} & 1 & \overline{\zeta_0}\zeta_1 &  & 0 & 0 \\
0 & 0 & \zeta_0\overline{\zeta_1} & 1 & \ddots  &  & 0 \\
\vdots  & \vdots  & & \ddots & \ddots & \overline{\zeta_0}\zeta_1 & \vdots\\
0 & 0 & 0 &  &  \zeta_0\overline{\zeta_1}  & 1 & \overline{\zeta_0}\zeta_1 \\
0 & 0 & 0 & 0 & \cdots & \zeta_0\overline{\zeta_1} & |\zeta_1|^2
\end{array}
\right)
\end{eqnarray*}
Let us note that 
$M=M_{m+1,1}(1,-|\zeta_1|^2,-|\zeta_0|^2,z,\overline{z})$, (see \cite{losonczi}) and $z=\overline{\zeta_0}\zeta_1$. 
In order to apply Theorem \ref{thm1} one should calculate $\Vert M\Vert_{(k)}$.
If $z=0$ then $M$ is a projection with $\Tr (M)=m$, so $\Vert M\Vert_{(k)}=k$ for $k=1,2,\ldots,m$.
Assume $z\neq 0$. Observe that 
\begin{equation}M-\lambda\mathds{1}=M_{m+1,1}(1-\lambda,-|\zeta_1|^2,-|\zeta_0|^2,z,\overline{z}).\end{equation} 
Therefore, to find eigenvalues of $M$ one needs to solve the equation
\begin{equation}
|z|^{n}\left(U_{m+1}\left(x\right)-\frac{1}{|z|}U_{m}\left(x\right)+U_{m-1}\left(x\right)\right)=0,
\label{eqn:det}
\end{equation} 
where 
$U_j$ are Chebychev polynomials of the second kind, and 
\begin{equation}x=\frac{1-\lambda}{2|z|}.\end{equation}
Chebyshev polynomials satisfy the following recurrence relations
\begin{equation}
\begin{split}
& U_{-1}(x)=0,\qquad U_0(x)=1  \\
& U_{m+1}(x)+U_{m-1}(x)=2xU_{m}(x), \quad m=0,1,\ldots. 
\end{split}
\end{equation}
Thus the equation \eqref{eqn:det} is equivalent to 
\begin{equation}
\label{eqn:determinant}
-\lambda |z|^{m}U_{m}\left(\frac{1-\lambda}{2|z|}\right)=0.
\end{equation}
It follows that roots of $U_m$ are precisely the numbers $\cos\left({j\pi}/(m+1)\right)$, for $j=1,2,\ldots,m$. Thus, the decreasing sequence $\lambda_1>\lambda_2>\ldots>\lambda_{m}$ of non-zero eigenvalues of $M$ is given by
$$
\lambda_j=1+2|z|\cos\left(\frac{j\pi}{m+1}\right),\quad j=1,\ldots,m,
$$ 
so,
\begin{equation}
\Vert M\Vert_{(k)}=k+2|z|\sum_{j=1}^k\cos\left(\frac{j\pi}{m+1}\right),\quad k=1,2\ldots,m.
\end{equation}
Taking into account that $\sup\{\left|\overline{\zeta_0}\zeta_1\right|\colon\Vert\zeta\Vert=1\}=\frac{1}{2}$, we obtain \eqref{eqn:thm6}.
\end{proof}

\begin{exam}
\label{prz}
Consider the case $m=3$ and $n=4$. In virtue of the Theorem \ref{thm:r_1}, the map $\Phi_a$ is positive for $a\geq 1+\frac{\sqrt{2}}{2}$, it is 2-positive for $a\geq 2+\frac{\sqrt{2}}{2}$, while it is completely positive for $a\geq 3$. 
\end{exam}

\begin{exam}
In \cite{chk} a technique based on a family of spectral conditions is developed to partially solve the problem of classification of positive maps. As an application of the main result of the paper some sufficient conditions on $k$-positivity for maps of the form \eqref{eqn:1} are given. However, for the case $m=3$, $n=4$, the criterion in \cite{chk} shows no range of the parameter $a$ for which the map \eqref{eqn:1} is $2$-positive and not completely positive. Namely, observe that the map \eqref{eqn:1} for $m=3$ and $n=4$ considered in the previous example have the Choi matrix
\begin{equation}\label{eqn:proposed_chk_form}
    C_\phi = a\mathds{1} - P_1 - P_2
\end{equation}
where $P_\alpha =\ket{\xi_\alpha}\bra{\xi_\alpha} \in \mathbb{M}_3\otimes\mathbb{M}_4$,
$\ket{\xi_\alpha}=\frac{1}{\sqrt{3}}\sum_{i=1}^3\ket{e_i}\otimes\ket{f_{i+\alpha}}$, $\alpha=0,1$. 
If $p\in\mathbb{M}_4$ is a projector with $\Tr(p)=2$ then $\norm{(\mathds{1}\otimes p)P_\alpha(\mathds{1}\otimes p)}=\frac{2}{3}$, for $\alpha=1,2$. Consequently $\sum_{\alpha=0}^1\norm{P_\alpha}_{(2)}^2=\frac{4}{3}$.
Therefore, the criterion from \cite[Theorem 2]{chk} is not applicable for the map \eqref{eqn:1}.
\end{exam}

\section*{Acknowledgments}
We would like to thank Adam Rutkowski for the fruitful discussions.
HO is supported in part by an KAKENHI Grant Number JP20K03644. 
TM is a PhD student at the Doctoral School of Natural Sciences of the University of Gda\'nsk.

\bibliography{newnew}
\bibliographystyle{abbrv}
\end{document}